\documentclass[12pt]{iopart} 
\expandafter\let\csname equation*\endcsname\relax

\expandafter\let\csname endequation*\endcsname\relax

\usepackage[hyphens]{url}
\usepackage{graphicx}
\usepackage{enumitem}
\usepackage{amsmath}
\usepackage{amsfonts}
\usepackage{amssymb}
\usepackage{color}
\usepackage{flushend}
\usepackage{soul}

\newcommand{\beq}{\begin{equation}}
\newcommand{\eeq}{\end{equation}}

\newcommand{\ketm}{\rvert -\rangle}
\newcommand{\ketp}{\rvert +\rangle}

\newcommand{\ngg}{n_\textrm{g}}

\newcommand{\Gp}{\Gamma}
\newcommand{\GG}{\gamma_\textrm{1p}}
\newcommand{\tp}{\tau_\textrm{p}}
\newcommand{\ts}{\tau_\textrm{s}}

\newcommand{\DL}{D$\Lambda$}

\newcommand{\Pc}{P_\textrm{c}}

\begin{document}

\title{Continuous generation of delayed light}

\author{Slava Smartsev$^1$, David Eger$^1$, Nir Davidson$^1$, and Ofer Firstenberg$^1$}

\address{$^1$ Department of Physics of Complex Systems, Weizmann Institute of Science, Rehovot 76100, Israel}
\ead{slava.smartsev@weizmann.ac.il}
\begin{indented}
\item[]June 2017
\end{indented}

\begin{abstract}
We use a Raman four-wave mixing process to read-out light from atomic coherence which is continuously written.
The light is continuously generated after an effective delay, allowing the atomic coherence to evolve during the process.
Contrary to slow-light delay, which depends on the medium optical depth, here the generation delay is determined solely by the intensive properties of the system, approaching the atomic coherence lifetime at the weak driving limit.
The generated light is background free. We experimentally probe these properties utilizing spatial diffusion as an 'internal clock' for the atomic evolution time. Continuous generation of light with a long intrinsic delay can replace discrete write-read procedures when the atomic evolution is the subject of interest.
\end{abstract}

%
\section{Introduction}

The interplay between optical fields and long-lived coherences in multi-level atoms provides a rich playground for fundamental and practical research. The basic process involves two fields and three levels in a $\Lambda$ or ladder configuration, giving rise to Raman transitions, electromagnetically-induced transparency (EIT) \cite{FleischhauerRMP2005}, and coherent population trapping (CPT) \cite{Arimondo96}, and leading to implementations of frequency standards \cite{knappe:1460}, magnetometers \cite{BudkerRomalisNP2007}, 
quantum memories and sources \cite{EisamanNature2005,KuzmichNatPhys2008,WalmsleyNJP2015},
and quantum nonlinear optics \cite{FirstenbergJPB2016,FirstenbergNature2013,HofferberthPRL2014,DuerrScience2016}.
In applications relying on EIT and CPT, the contrast of the transmitted signal and the group delay in the medium play a crucial role, determining the metrological accuracy or the process fidelity.

From the basic tools of EIT and CPT, one can construct composite schemes, involving 3 or 4 fields and 4 levels \cite{LukinAdvAtMolOptPhys2000}. Various 4-wave mixing processes have been investigated, particularly for quantum optics, such as generation of entangled and squeezed light \cite{LettNature2009,PolzikRMP2010} and single-photon sources \cite{MatsukevichPRL2006}. Here, we study a distinct 4-wave mixing regime, where a signal field is generated with near unity contrast after a finite delay time. As opposed to standard slow light via EIT, here the signal delay is determined only by the linewidth of the Raman transition and is insensitive to extensive variables (\emph{e.g.}, number of atoms, population distribution, optical depth). The light is continuously generated from a Raman coherence that has evolved for a finite duration. 


Let us briefly review the various existing schemes involving two strong \emph{control} fields and two weak \emph{probe} fields. Parametric gain, via the conversion of two control photons to two probe photons, occurs in a so-called \emph{closed} double-$\Lambda$ (\DL) configuration and leads to entangled (twin) photon generation \cite{PolzikRMP2010,GlorieuxPRA2010}. Parametric conversion, where the two probe fields are effectively coupled, and the total number of probe photons is conserved, occurs in an \emph{open} \DL~\cite{MandelPRA1990} and a double-V (DV) \cite{ArimondoOptComm1996} configurations. Our configuration is a DV, which is the least studied \cite{ArimondoOptComm1996,GaoOE2011,KaniOE2014}.

In an open \DL, the control fields form a $\Lambda$ system with two ground-states and one excited state. This leads to CPT with a Raman coherence that is generally a spatial grating, thus the name `electromagnetically induced grating' \cite{XiaoPRA1998}. Due to motion of the atoms through the grating, the process is sensitive to the wave-vector mismatch between the two control fields.
The probe fields form a second $\Lambda$-system with a different excited state and experience the induced grating \cite{ArimondoPRA1995,KosachiovPRA1999}. For a single incoming probe, its diffraction by the grating generates the second probe \cite{HemmerOL1995}, providing a Raman-spectroscopy signal with near-unity contrast \cite{KitchingOL2007}. However the probe fields are not substantially slowed, as the control fields alone dictate the Raman detuning.

\begin{figure}[tb]
    \includegraphics[width=7.8cm,trim={0.3cm 4.5cm 0cm 0cm},clip]{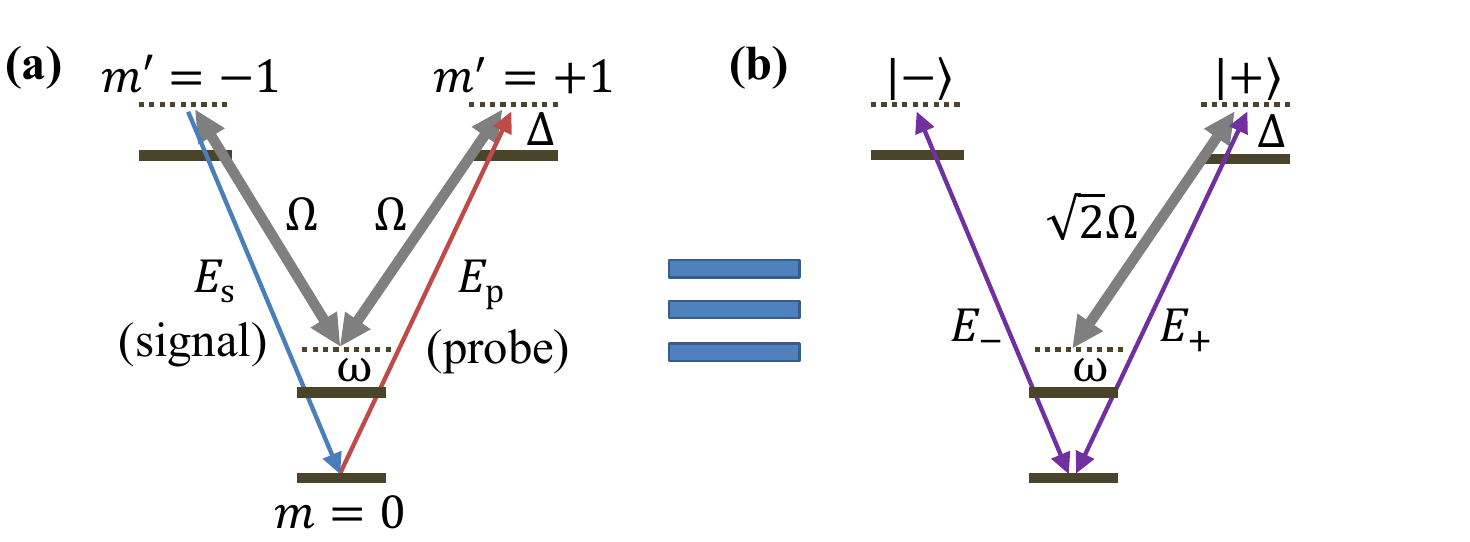}
	\includegraphics[width=7.8cm,trim={0.3cm -0.85cm 0cm 5.7cm},clip]{Figure_30.pdf}
	\caption{(a) Double-V configuration comprising two strong control fields (thick gray) and weak probe (red) and signal (blue) fields. (b) In the rotated basis $|\pm\rangle=(|m'=-1\rangle\pm|m'=1\rangle)/\sqrt{2}$, only one of the excited states couples to the control fields. The superposition $E_+$ of the probe and signal fields experiences EIT, while $E_-$ does not. (c) The probe is transmitted, and the signal is generated. (d) An interferometer interpretation:  In the medium, the probe is split into $E_+$ and $E_-$ with corresponding transmission amplitudes $g_+$ (with EIT) and $g_-$ (without EIT). The difference $g_+-g_-$ leads to constructive interference at the signal output.}
	\label{fig_scheme}
\end{figure}

In a DV configuration, depicted in Fig.~\ref{fig_scheme}(a), the control fields couple one ground state to two excited states. No ground-state coherence is formed without the probe fields, and thus no grating is created. Similarly to the open \DL, one incoming probe field generates the other;
However, in the DV, the process is sensitive to the Raman detuning between the probe and the control and insensitive to that between the controls. The scheme we study utilizes both these properties.

Note that the DV configuration is equivalent to the double ladder configuration \cite{WelchOL2008} --- where control (probe) fields drive the upper (lower) transitions --- but the Raman coherence is usually longer lived in the DV. The efficiency of 4-wave mixing in both configurations has received considerable attention \cite{WelchOL2008,OrozcoPRA2008,GaoOE2011}. Here we focus on the temporal properties of the process. Contrary to a recent work \cite{BaoSenChinPhysB2013}, we study the regime of very long intrinsic delay times, manifesting the continuous read-out of light.

\section{Experiment}
We use $^{87}$Rb vapor and 10 Torr of N$_2$ buffer gas in a cell of length 7.5 cm at $55^\circ$C. The DV system [Fig.~\ref{fig_scheme}(a)] comprises the $|F=1,2;m=0\rangle$ states of the ground 5S$_{1/2}$ level and the $|F'=1,2;m=\pm1\rangle$ states of the excited 5P$_{1/2}$ level of the D1 transition. We define the one-photon detuning $\Delta$ from $F'=2$ and experimentally vary $\Delta$ over a $\sim$1 GHz range, such that generally both $F'=1$ and $F'=2$ participate in the process.
A 50-mG longitudinal magnetic field guarantees that the spectator ground states $|F=1,2;m\ne0\rangle$ are far from the Raman resonance conditions. The light from an amplified 795-nm diode laser is split into one probe and two control beams. We shift the probe frequency by the hyperfine splitting $\sim$6.8 GHz using optical modulators, followed by an etalon filter. These are also used to scan the two-photon detuning $\omega$ and temporally shape the probe pulses. 
The probe 
is aligned with one control beam,
while the second control beam is sent at a mechanically-controlled angular deviation $\theta=5-27$ mrad. All beams are collimated to a diameter $\sim$8 mm and spatially overlap when crossing the vapor cell. After the cell, the probe and signal angularly separate, and we filter them from the control light using etalons. The resonant optical depth $2d$ for the probe and signal increases from $2d=5$ at a control power $\Pc=.01$~mW to $2d=10$ at $\Pc=2$~mW.

\begin{figure}[tb]
	\includegraphics[width=7.7cm,trim={0cm 4.5cm 0cm 0cm},clip]{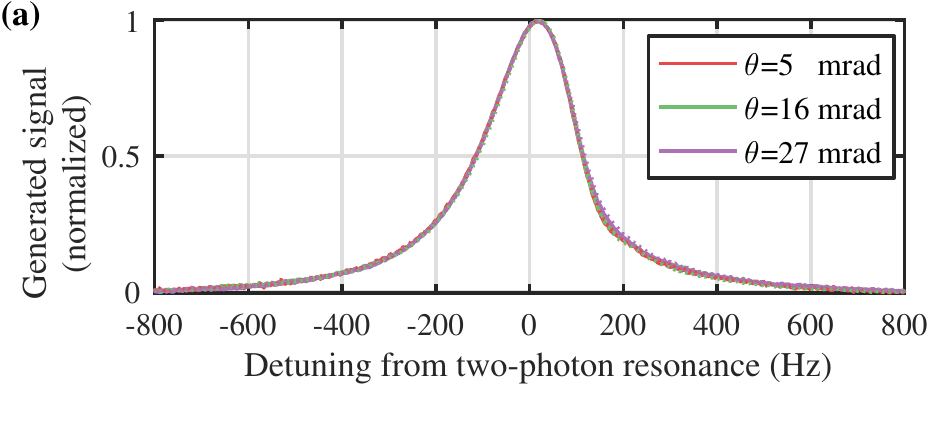}
	\includegraphics[width=7.7cm,trim={0cm 0cm 0cm 4.5cm},clip]{Figure_40.pdf}
	\caption{(a) Generation spectra of the signal for different angular deviations $\theta$ with the circular polarizations of the two control beams either (solid lines) perpendicular or (dashed lines) parallel; $\Pc=0.14$~mW. 
		The six normalized spectral lines of width $\sim 230$~Hz are nearly identical, demonstrating the insensitivity of the process to polarizations and to $\theta$.
		(b) Contrast of the spectral lines of the signal and probe fields versus $\Pc$. 
		The contrast is defined as $|P_\textrm{r}-P_\infty|/|P_\textrm{r}+P_\infty|$, where $P_\textrm{r}$ ($P_\infty$) is the measured probe or signal power on (off) the two-photon resonance. With decreasing $\Pc$, the contrast of the probe drops dramatically, whereas that of the signal remains high (eventually, the signal contrast drops when the generated power reaches the noise level off resonance). 
	}
	\label{fig_spectra}
\end{figure}

%
The robustness of the scheme is demonstrated by the generation spectra in Fig.~\ref{fig_spectra}(a). The lineshape does not depend on the relative polarizations of the control beams or on the deviation angle between them (for $\theta\ll 1$). 
This confirms that no spatial grating of the ground-state coherence is formed by the controls. In particular, the phase-matching conditions are equally fulfilled regardless of $\theta$, since $\theta$ does not impact each of the two $\Lambda$ systems separately.
Note that this property is exploited in counter-propagating DV configurations ($\theta=180^\circ$) \cite{LukinNature2003,YuPRL2009}.


The measurements in Fig.~\ref{fig_spectra}(a) are done at $\Delta=880$~MHz, where the \emph{transmission} spectrum has a characteristic Fano-like lineshape. However, the  \emph{generation} lineshape remains a Lorentzian regardless of $\Delta$, further manifesting the robustness of the lineshape.

Figure \ref{fig_spectra}(b) presents a comparison between the contrasts of the generation and transmission spectra. 
While the contrast of the transmitted probe vanishes for low control power, the signal is generated with no background and thus with near unity contrast regardless of the control parameters.

\begin{figure}[tb]
	\includegraphics[width=7.7cm,trim={0cm 4.55cm 0cm 0cm},clip]{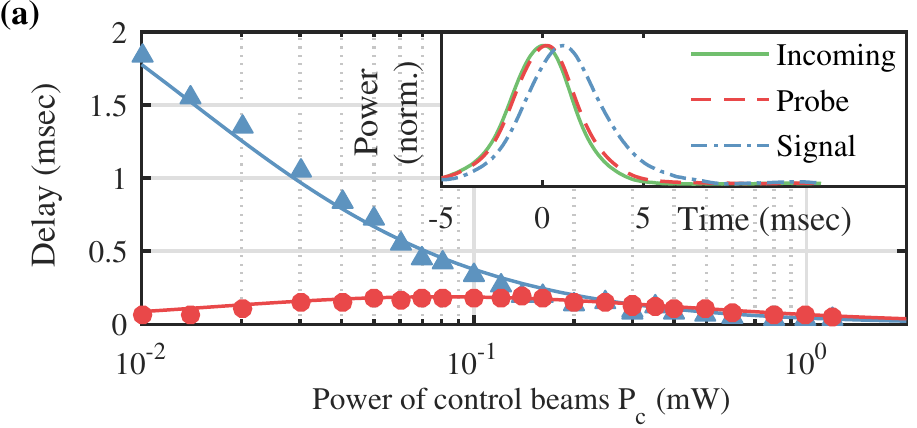}
	\includegraphics[width=7.7cm,trim={0cm -0.08cm 0cm 4.5cm},clip]{Figure_10.pdf}
	\caption{(a)
		Measurements of pulse delay at $\Delta=0$.
		The signal delay $\ts$ (blue) and probe delay $\tp$ (red) differ most substantially at low control power: $\ts$ depends only on the inverse linewidth and thus increases at low power, while $\tp$ diminishes due to loss of contrast of the transmission line
		(inset shows a measurement example for $\Pc=0.03$~mW).
		The lines are model fits.
		(b) $\tp$ is highly sensitive to $\Delta$ compared to $\ts$, exemplifying the robustness of the generation properties. 
	}
	\label{fig_delay}
\end{figure}


We send probe pulses, shown in Fig.~\ref{fig_delay} (inset), to measure the delay of the transmitted probe $\tp$ and the delay of the generated signal $\ts$. As summarized in Fig.~\ref{fig_delay}(a),
$\tp$ vanishes when the control power is lowered, 
whereas $\ts$ monotonically increases. 
This striking feature of the generation process stems from its near-perfect contrast [Fig.~\ref{fig_spectra}(b)], as the group delay is proportional to the ratio between the contrast and width of the spectral line \cite{WalsworthPRA2009}. While this ratio renders a well-known tradeoff in the probe channel (both contrast and width increase with control power), it is clearly maximized for weak control in the signal channel.
Fig.~\ref{fig_delay}(b) shows that for low control power, $\ts$ is insensitive to the one-photon detuning $\Delta$. In contrast, $\tp$ decreases at large detuning $|\Delta|$.

As we will show, $\ts$ approaches the lifetime of the Raman coherence $\gamma^{-1}$ (measured 3 ms in our system) for low control power.
Indeed this intuitively should be the upper limit: The probe and the co-propagating control act to `write' the ground-state coherence, having a lifetime $\gamma^{-1}$, and the second control `reads' it. This implies that the signal is continuously generated from atomic Raman coherence that has evolved for a duration $\ts$.


To validate this picture, we measure the spatial evolution in the planes transverse to the propagation direction. 
The transverse evolution 
in the paraxial regime are largely separable from the longitudinal evolution \cite{FirstenbergPRA2008}, providing an effective internal clock for the continuous generation process. In standard 'light storage' procedures, the ground-state coherence of the diffusing atoms undergoes spatial diffusion \cite{ShukerImaging2007,FirstenbergPRL2010}. Consequently for our process, the transverse distance traveled by the diffusing atoms from which the signal light is read quantifies the effective atomic evolution time.

\begin{figure}[tb]
	\centering
	\includegraphics[width=8.5cm]{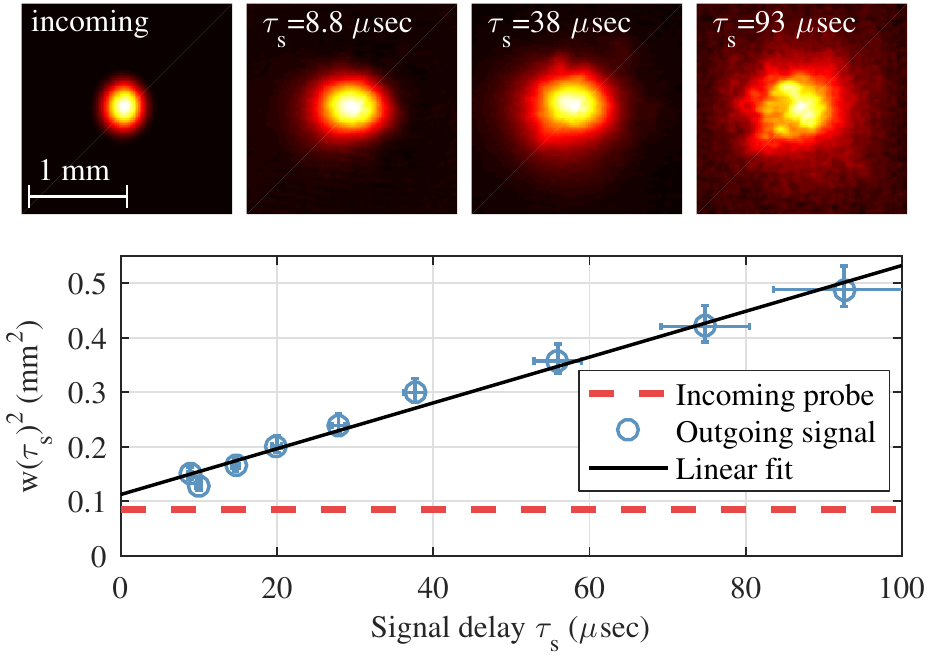}
	\caption{Diffusion of a Gaussian field during the process. Top: Incoming probe beam compared to generated signal beams for three signal delays $\ts$.
		Bottom: Measured area of the signal beam versus $\ts$. The linear fit agrees with the expected diffusion broadening. The overlap of the measured profile with an ideal Gaussian intensity distribution is $>90\%$ for all these data.}
	\label{fig_diffusion}
\end{figure}

We perform experiments with a narrow Gaussian probe beam and image the generated signal onto a camera (Fig.~\ref{fig_diffusion}). We vary the signal delay $\ts$ by altering the control power $\Pc$. The one-photon detuning is $\Delta\approx700$~MHz; the small shift of the $|F=1\rangle$ state due to the varying power of the off-resonant control is compensated for by fine tuning the probe frequency, thus maintaining the two-photon resonance. The Gaussian shape is imprinted on the atoms and expands due to atomic diffusion. Using the measured Gaussian width $w(\ts)$, we fit $w(\ts)^2=w(0)^2+4D\ts$ \cite{FirstenbergPRL2010} to extract the diffusion coefficient $D\approx1050$~mm$^2$/s. This matches the calculated value \cite{YabuzakiPRA2000,HapperPRA2009}, which we also measure in independent storage experiments. We conclude that the atomic coherence evolves for the duration $\ts$ before generating the signal.

\section{Theoretical model}

Full Maxwell-Bloch models of DV systems are known and have been solved for various scenarios \cite{ArimondoOptComm1996,KaniOE2014,WelchOL2008,OrozcoPRA2008,BaoSenChinPhysB2013}. Here to construct a simple model, we notice that the control beams couple the ground state to a specific superposition of the two excited states, denoted as $\ketp$ in Fig.~\ref{fig_scheme}(c). 
Thus instead of the bare probe and signal fields, it is convenient to use the normal modes $E_+$ and $E_-$, which respectively couple to $\ketp$ and its orthogonal superposition $\ketm$. Here, the $E$'s denote the slowly varying field envelopes in space and time. Only $E_+$ experiences EIT (note the difference with the open \DL~scheme, where neither of the normal modes experiences EIT \cite{LukinAdvAtMolOptPhys2000}).
The medium then acts as the effective Mach-Zehnder interferometer in Fig.~\ref{fig_scheme}(d), where the incoming probe decomposes into $E_+^{\textrm{in}}$ and $E_-^{\textrm{in}}$ upon entering the medium.


Instead of solving coupled propagation equations for the bare modes, we immediately express the normal modes at the medium exit as $E_\pm^{\textrm{out}}=2g_\pm E_\pm^{\textrm{in}}$ using the known linear susceptibility with and without EIT. The transmission amplitudes with and without EIT are, respectively, $2g_+=e^{-S (1-f)}$ and $2g_-=e^{-S}$ \cite{FleischhauerRMP2005}, where
\beq
S=d\frac{\GG}{\GG-i\Delta} \textrm{~~and~~} f=\eta_{\textrm{act}}\frac{\Gp}{\gamma+\Gp-i\omega} \label{eq_Sf}
\eeq
are complex Lorentzians associated with the one-photon and two-photon resonances, with $\GG$ and $\gamma$ the corresponding decoherence rates. 
Both the power broadening term $\Gp=2\Omega^2/(\GG-i\Delta)$ and the ratio of `active' atoms (in the $\Lambda$-system) to `spectator' atoms $\eta_{\textrm{act}}$
depend on the Rabi frequency $\Omega$ of the control beams (assumed equal). The real and imaginary parts of $\Gp$ correspond respectively to line broadening and to light shift at $\Delta\ne 0$.
Note that the model, here presented for atoms at rest, can be generalized for a thermal medium 
predominantly by correcting $S$ and $\Gp$ to account for the broadening of the 
one-photon spectrum \cite{FirstenbergPRA2008}.


For an incoming probe field $E_\textrm{in}$, the transmitted probe is given by $E_\textrm{p}=E_\textrm{in}(g_++g_-)$ and the generated signal by $E_\textrm{s}=E_\textrm{in}(g_+-g_-)$.
A signal is generated when $g_+\ne g_-$, \emph{i.e.}~when $|S f|\ne 0$ due to EIT.
We identify the limit of weak EIT with $|S f(\omega=0)|\ll 1$, which is oftentimes the case even in optically thick media: Limited control power and a desire to minimize power broadening restrict the induced transparency. Additionally, `spectator' atoms residing in states outside the $\Lambda$-system (here $|F=1;m\ne0\rangle$) 
contribute to the absorption but are unaffected by EIT ($\eta_{\textrm{act}}<1$).

The generation efficiency $\beta=|E_s/E_\textrm{in}|^2$ at the weak EIT limit $|S f|\ll 1$ is low, as can be calculated from
\beq
E_s/E_\textrm{in}=g_+-g_-=e^{-S}(e^{S f}-1)/2\approx e^{-S}S f/2. \label{eq_t}
\eeq
For the experimental conditions of Figs.~\ref{fig_spectra} and \ref{fig_delay}(a), the measured efficiency at low control power is $\beta=10^{-4}\sim 10^{-3}$, agreeing with that estimated from the above expression ($S = d \sim 2.5$ and $f \gtrsim 0.1$). We shall assume that $S$ and $\Gp$ vary only slightly with $\Delta$ near the EIT resonance. This assumption holds for an EIT resonance much narrower than the optical resonance, as occurring in a hot vapor. The generation spectrum near resonance thus has the shape $\beta\propto |f(\omega)|^2$. Notably, it is Lorentzian regardless of $\Delta$ as evident by Fig.~\ref{fig_spectra}(b), as opposed to the asymmetric (Fano-like) lineshape of the transmitted probe at large $\Delta$.


We now turn to explain the striking features of the generated signal observed in the temporal and spatial domains.
The delay of the transmitted probe $\tp$ stems from a well-studied mechanism of slow light due to the large group index $\ngg\gg 1$ on EIT resonance. When $\ngg\gg 1$, the group delay of the probe and signal can be calculated for the whole medium using $\tau=\textrm{Im}[\frac{\partial}{\partial\omega}\ln(g_+ \pm g_-)]$
(This expression is easily understood for a purely dispersive medium by replacing $g_+ \pm g_-$ with $e^{-i\omega n_g L/c}$ to find $\tau=n_g L/c$).
From $g_+ +g_-= e^{-S}(1+e^{S f})/2$ and Eq.~(\ref{eq_Sf}), and assuming $|S f|\ll 1$ and negligible $\partial S/\partial \omega$ and $\partial \Gp/\partial \omega$,
we find the group delay of the probe
\beq
\tp=\textrm{Re}\frac{S}{2}\frac{\eta_{\textrm{act}}\Gp}{(\gamma+\Gp)^2}
\label{eq_tau_p}
\eeq
at $\omega=0$.
For $\Delta=0$, all these parameters are real, and $S=d$. The dependence of $\tp$ on the optical depth (and thus on the atomic density and the length of medium), as well as on the population ratio $\eta_\textrm{act}$
and on the control power (via $\Gp$), 
is well-known for EIT slow-light \cite{WalsworthPRA2009}.

With Eqs.~(\ref{eq_Sf}) and (\ref{eq_t}), we find the group delay of the signal at $\omega=0$
\beq
\ts=\textrm{Re}\frac{1}{\gamma+\Gp}. \label{eq_tau_s}
\eeq
As opposed to standard `delay lines', and in contrast to the probe delay, the signal delay is independent of any extensive parameter, such as the medium length, optical depth, or level population. It is also independent of the EIT contrast, remaining finite for weak control fields. For vanishing control $\Gp\rightarrow 0$, the probe delay vanishes $\tp\rightarrow 0$, while the signal exhibits maximal delay $\ts\rightarrow \gamma^{-1}$ as observed in Fig.~\ref{fig_delay}(a).

For fitting $\ts$ in Fig.~\ref{fig_delay}(a), we use $\ts=[\gamma+\Gp(\Pc)]^{-1}$, were the power broadening $\Gp(\Pc)$ versus $\Pc$ is calibrated from independent spectra measurements for a range of $\Pc$. The value $\gamma=$(3 msec)$^{-1}$ is found by linear extrapolation of $\ts^{-1}$ to $\Pc\rightarrow 0$. $\tp$ is described by Eq.~(\ref{eq_tau_p}), 
depending additionally on $S\eta_\textrm{act}$.
The optical depth and population distribution vary due to optical pumping, and thus $S\eta_\textrm{act}$ depends on $\Pc$. We calibrate this dependence from measured spectra.
The need to calibrate the optical pumping processes in standard slow-light applications exemplifies the intricate dependence of $\tp$ on various system parameters versus the 'naturalness' of $\ts$.


To describe spatially-structured fields, we work in Fourier space of the transverse planes, 
where the field envelopes of the probe and signal are functions of the transverse wave-vector $\mathbf{k}$. The Raman process depends on $k\equiv|\mathbf{k}|$, since nonzero $k$ represents a wave-vector mismatch between the probe or signal and their corresponding control field.
The $k$-dependence of the process is a consequence of motional line broadening. The diffusion of Rb in the buffer gas across the Raman wavelengths ($\propto k^{-1}$) results in a broadening quadratic in $k$ due to the Dicke effect. 
To account for this, one only needs to generalize $f$ in Eq.~(\ref{eq_Sf}) to a $k$-dependent $f^{(k)}$ by replacing $\gamma\rightarrow\gamma+Dk^2$ \cite{FirstenbergPRA2008,FirstenbergRMP2013}. In this description, we ignore paraxial diffraction, the $\mathbf{k}$ mismatch due to the hyperfine splitting, and a small offset to $\mathbf{k}$ (on order $\propto \theta^2$) due to the angular deviation of the controls; all of these can easily be added to the model and are unimportant at our conditions.

It has been shown previously for slow light \cite{FirstenbergNatPhys2009} that the quadratic broadening $Dk^2$ results in diffusion of the probe envelope for the duration of the group delay $\tp$, under the conditions of resonant excitation ($\Delta=\omega=0$) and confined spatial spectrum $k^2\ll|\gamma+\Gp|/D$.
Under these conditions,
\beq
f^{(k)}=\frac{\gamma+\Gp}{\gamma+\Gp+D k^2}f^{(k=0)} \approx e^{-\ts Dk^2}f^{(k=0)}.
\eeq
Now for the signal field, we substitute $f^{(k)}$ into Eq.~(\ref{eq_t}) and find $E_\textrm{s}^{(k)}\propto E_\textrm{in}^{(k)} e^{-\ts Dk^2}$. 
The `filter' $e^{-\ts Dk^2}$ in $k$-space leads to diffusion in real space for the duration $\ts$.
This solution agrees with the physical picture of light generation after an atomic evolution time $\ts$.
Figure \ref{fig_diffusion} 
demonstrates the validity of this picture, even beyond the weak EIT condition (with $S\eta_\textrm{act}\approx 2$).

\section{Conclusions}
In conclusion, the continuous generation process 
facilitates studies of atomic evolution (internal or external) and interatomic interactions. The absence of background incident light in the outgoing signal offers reduced noise for metrology applications and higher purity for quantum-optics applications.
One exciting prospect is to utilize the atomic evolution for generation of non-classical signal fields, particularly anti-bunched light. Generation of anti-bunched light due to interactions between Rydberg atoms was recently demonstrated in a discrete write-read procedure \cite{KuzmichScience2012}. The same mechanisms, applied for the effective duration of the atomic evolution in our scheme, could provide a continuous source of single photons. Here the low generation efficiency does not limit the anti-bunching fidelity. Furthermore in contrast to continuous collinear configurations, such as Rydberg-EIT \cite{PeyronelNature2012}, the anti-bunching statistics in our scheme will not be limited by the optical depth. Our scheme is thus particularly suitable for media with low optical depth, \emph{e.g.}~in miniature hot-vapor cells.

\section*{Acknowledgments}
We acknowledge financial support by the Israel Science Foundation and ICORE, the Laboratory in Memory of Leon and Blacky Broder, and the Sir Charles Clore research prize.

\section*{References}

\bibliographystyle{iopart-num} 
\bibliography{20170314_Generation_bib}

\end{document}